\documentclass[aps,prb,twocolumn,superscriptaddress,groupedaddress]{revtex4}
\usepackage{graphicx}
\usepackage{dcolumn}
\usepackage{bm}       
\usepackage{amssymb} 
\usepackage{amsmath}
\usepackage{amsfonts}
\usepackage{amsthm}
\usepackage{xfrac}
\usepackage[cp1252]{inputenc}
\usepackage{epsf}
\usepackage{lpic}
\usepackage{tikz}
\usepackage{circuitikz}
\usepackage{physics}
\usepackage{mathtools}
\usepackage{subcaption}
\newcommand{\be}{\begin{equation}}
\newcommand{\ee}{\end{equation}}
\newcommand{\ba}{\begin{align}}
\newcommand{\eal}{\end{align}}
\newcommand{\dg}{ ^{\dagger}}

\begin{document}

\title{Realization of the degenerate parametric oscillator in electromechanical systems}
\author{E. Jansen}
\author{J. D. P. Machado}
\author{Ya. M. Blanter}
\affiliation{Kavli Institute of Nanoscience, Delft University of Technology, Lorentzweg 1, 2628 CJ Delft, The Netherlands}

\date{\today}

\begin{abstract}
We consider an electromechanical system where a microwave cavity is coupled to a mechanical resonator, with a mechanical frequency twice the microwave frequency. In this regime, the effective photon-phonon interaction is equivalent to that of a degenerate parametric amplifier, instead of the typical radiation pressure interaction. If the mechanical resonator is strongly driven, it undergoes a phase transition to a state where the energy pumped into the mechanical mode is entirely converted to the photonic mode. Quantum fluctuations smear this phase transition. We describe these effects with a steady-state Fokker-Planck equation in the complex P-representation, and compute the photonic field intensity and quadrature variances, as well as the mechanical amplitude. This Fokker-Planck method performs better than the standard linearization results, when compared to numerical simulations.
\end{abstract}

\maketitle

\section{Introduction}
The field of optomechanics, which studies the interaction of light with mechanical motion in an optical cavity, has been rapidly developing in the last decade \cite{ Aspelmeyer}. It was initiated because of the interest in the quantum measurement limit in gravitational wave detectors \cite{braginsky}. More recently, interest has risen in the utilization of opto- and electromechanical systems in quantum information processing \cite{Stannigel, Marquardt} and for fundamental tests of quantum physics \cite{Penrose}. 

In parallel, research has been performed on mechanical resonators coupled to superconducting microwave circuits. This field is usually referred to as microwave optomechanics, since, despite big differences in setup and in parameter regimes, the physics these systems display is remarkably similar to that of optomechanical systems. In particular, similarly to optomechanics, mechanical resonators couple to microwave radiation via radiation pressure. Indeed, the simplest realization of such a system is a LC-circuit where one of the capacitor's plates is a vibrating drum (capacitive coupling). Typically, the LC resonance frequency ($\omega_{\text{LC}}\sim \text{GHz}$) is much bigger than the mechanical frequency $(\Omega_{\text{m}}\sim \text{MHz})$, leading to the usual radiation pressure Hamiltonian \cite{Aspelmeyer}. Similarly, an inductive coupling, where a mechanical element forms a movable part of a superconducting loop, leads to radiation-pressure interaction \cite{Shevchuk}. Experiments with microwave optomechanics demonstrated a number of remarkable effects, including sideband cooling \cite{Teufel2011}, reaching the quantum ground state of a mechanical oscillator and controllably creating single phonon excitations \cite{Connell}, and measurement of mechanical motion near the quantum limit \cite{Teufel2009}.

When the frequency of the mechanical resonator is much lower than the cavity frequency, the effective interaction takes the usual form \cite{Aspelmeyer} $a\dg a (b\dg+b)$, but different forms of interaction can be attained for higher mechanical frequencies. Mechanical resonators in the GHz frequency range exist and have been integrated into optomechanical systems (see {\em e.g.} Refs. \onlinecite{Chan2011, Lin2011, Sun2011, Pfeifer2016}), and a relatively large coupling rate of $g \sim 10^5$ Hz has already been reported in the microwave regime \cite{Schoelkopf2017}. Alternatively, in principle, the LC resonance frequency can be lowered to the MHz-level by using a large inductance or capacitance.

Microwave circuits with integrated mechanical resonators (to be referred below as {\em electromechanical systems}) in the regime when the microwave and mechanical frequencies have the same order of magnitude do not have the standard interaction term $a\dg a (b+b\dg)$ as the main coupling mechanism. Little is known about this regime, and investigation is needed to establish the basic behavior of such electromechanical systems. 

\begin{figure}[h]
\centering
\begin{tikzpicture}
--------start graphics code --------
\draw (0,0) node [ground] {}
			     -- (0,2)
  				 to [R=\(R\)] (5,2)
  				 to [L=\(L\)] (5,0) -- (3.1,0)
   				 to [spring=\(k\)] (2.35,0)
   				 to [C = \(C\)] (1.93,0) 
   				 -- (0,0);
\draw (4.65,-0.8) 
				to [sV] (3.8,-0.8)
				to (3.15,-0.8)
				to (3.15,0) ;
--------end graphics code ----------
\end{tikzpicture}
\caption{The lumped-element representation of a RLC-circuit where one of the capacitor plates is a vibrating drum. This plate is mechanically driven so that the effective distance $d_0 + x$ between the plates of the capacitor is varied. A resistor was included to depict the unavoidable dissipation affecting the circuit.}
\label{fig:fullcircuit}
\end{figure}
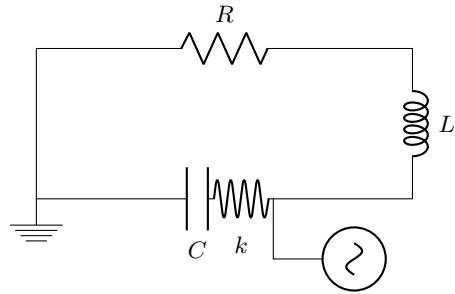

In this article, we study these systems in the {\em parametric regime}, defined by the condition $2 \omega_{\text{LC}} \sim \Omega_\text{m}$.
The circuit corresponding to the electromechanical system under consideration is shown schematically in Fig. \ref{fig:fullcircuit}. The effective Hamiltonian describing the system is of the form of the degenerate parametric oscillator, which is well known in the context of quantum optics \cite{Charmichaelbook1, MilburnWallsbook}. This Hamiltonian describes a special case of the parametric down-conversion process -- the phenomenon where an incoming pump photon with frequency $\Omega$ is converted to two photons with frequencies $\omega, \omega'$ such that $\Omega = \omega + \omega'$. This conversion can occur in a nonlinear medium like a nonlinear $\chi^{(2)}$-crystal \cite{GarrisonChiao}. Degenerate parametric down-conversion refers to the special case where $\omega = \omega'$. Classically, this system exhibits a phase transition at the critical driving strength. Below the transition, the amplitude of the pump (harmonic) mode increases linearly with the driving whereas the amplitude of the subharmonic mode equals zero. Above the transition, the pump amplitude becomes constant and all energy pumped into the system is converted to the subharmonic mode.

In electromechanical systems, if the mechanical element is driven, its motion modulates the capacitance (or inductance) of the circuit, and the down-conversion process corresponds to the creation of 2 microwave photons from a mechanical phonon. This photon creation process is connected to the Dynamical Casimir Effect (DCE) \cite{Mouro}, and the connection between the parametric oscillator and DCE connection has been explored in the optical domain \cite{Astrid}. Recently, there have been proposals to observe this effect in electromechanical systems \cite{simonetamigos}, valid below threshold. To enhance the creation of microwave photons, the mechanical resonator must be strongly driven, but over a certain driving strength, the backaction from microwave photons will affect the modulating mechanical element. Here, we take this backaction into account.

In the quantum optical case and in the usual regime of a fast-decaying pump mode, the dynamics is solved through adiabatic elimination of the pump mode \cite{Drummond1981, MilburnWallsbook}. However, in electromechanics the photonic (subharmonic) dissipation rate $\kappa$ is typically much bigger than the phononic (pump) dissipation rate $\Gamma$, so that the pump mode cannot be eliminated adiabatically. This regime is known as the {\em diabatic regime}. Furthermore, because the system undergoes a phase transition at the critical driving strength, a simple linearization procedure produces diverging results for the system's observables near the transition. To overcome this problem, a self-consistent linearization procedure was proposed \cite{Veits}. However, when both modes decay on the same timescale, the predictions of the self-consistent method deviate qualitatively from our numerical results. Here, we find the steady-state solutions for the system's observables by deriving an effective steady-state Fokker-Planck (FP) equation in the complex P-representation \cite{Drummondgen} equivalent to that describing the adiabatic regime \cite{Drummond1981} so that known results may be extended to the electromechanical situation. We argue that in the diabatic limit the fluctuations become small and this approach becomes a very good approximation. Away from the diabatic limit, it reduces to a linearization. We also compare the results of this approach to those of numerical simulations.

The paper is organized as follows. In Section II, we derive the Hamiltonian describing the circuit in Fig. \ref{fig:fullcircuit} and show that it has the form of the degenerate parametric oscillator. Then, we show how the phase transition occurs in the model and how self-consistently linearized solutions can be found. In Section III, we derive an effective FP equation that is equivalent to the FP equation describing the adiabatic regime and argue that it describes the steady-state of the system in the diabatic regime to a good approximation as well. We present analytical expressions for the photonic moments and mechanical amplitude by extending known results \cite{Drummond1981}. In Section IV, we compare the results of the FP method for the mechanical amplitude and the photon number with the results of the semiclassical (mean-field) approach, the self-consistent approach, and numerical simulations. These simulations show that above threshold, the system goes into a mixture of photonic coherent states. In Section V, we present the conclusions.

\section{The Hamiltonian}\label{Hamiltonian}
The circuit in Fig. \ref{fig:fullcircuit} describes two coupled harmonic oscillators: a LC resonator and a mechanical resonator. The energy stored in the system is given by
\be
H=\frac{1}{2L}\Phi^2+\frac{1}{2}C(x)Q^2+\frac{1}{2m}p^2+\frac{1}{2}m\Omega_m^2x^2\,,
\ee
where $\Phi$ and $L$ are respectively the magnetic flux through the inductor and the inductance, $Q$ the capacitor charge, $x$ and $p$ the displacement and momentum of the mechanical resonator, and $m$ its mass. For a parallel-plate capacitor, the capacitance varies with the position of the mechanical resonator as $C(x)=C_0\frac{d_0}{d_0+x}$, where $C_0$ and $d_0$ are the unperturbed capacitance and distance between the capacitor plates, respectively. With the quantization rules of quantum network theory \cite{quantumnetworktheory}, and including driving of the mechanical resonator, the quantum Hamiltonian corresponding to this circuit is
\begin{align}\label{eq:fullH}
H= &\omega_{LC} a\dg a + \Omega_m b\dg b - g \left(a\dg - a \right)^2\left(b\dg + b \right) \nonumber \\&+ \mathcal{E}\left(b\dg e^{-i\omega_D t} + b e^{i\omega_D t} \right),
\end{align}
where $a$ ($b$) is the annihilation operator for the LC (mechanical) resonator, $g = \omega_{LC} x_{\text{ZPF}}/(4d_0)$ is the single photon coupling rate, which scales linearly with the zero-point motion $x_{\text{ZPF}}$, and $\mathcal{E}$ is the driving amplitude. The explicit time dependence of the Hamiltonian can be removed by transforming to the rotating frame via the unitary transformation $U = \exp [-i \omega_D t (\frac{1}{2} a\dg a + b\dg b)]$. Driving the mechanical oscillator at its resonance frequency ($\omega_D = \Omega_m$), Eq.(\ref{eq:fullH}) becomes
\begin{align}\label{eq:fullH2}
H'&=UHU\dg +i(d_t U) U\dg= \nonumber\\
&=\big(\omega_{LC} -\frac{1}{2}\Omega\big) a\dg a  + \mathcal{E}\left(b\dg + b  \right)\nonumber\\
&- g \left(a\dg  e^{i\omega_{LC} t}- a  e^{-i\omega_{LC} t}\right)^2\left(b\dg e^{i\omega_D t} + b  e^{-i\omega_D t}\right) 
\end{align}
in the driving reference frame. When $\omega_{LC} \gg \Omega_m$, the rotating wave approximation (RWA) makes the effective interaction take the form of the usual radiation pressure term. Here, we are interested in the parametric frequency regime, where $2 \omega_{LC} = \Omega_m$. Invoking the RWA, the Hamiltonian (\ref{eq:fullH2}) reduces instead to the form of the degenerate parametric oscillator
\be\label{eq:paramH}
H = ig\left( a\dg a\dg b - a a b\dg \right) + i \mathcal{E} \left( b\dg - b \right),
\ee
where we rotated the $b$-field by a phase of $-\pi/2$ to make the resulting equations of motion real ($b\to -ib$). The Hamiltonian (\ref{eq:paramH}) describes the conversion of a phonon from the pump (phononic/harmonic) mode $b$ to two photons in the photonic (subharmonic) mode $a$. 
Taking dissipation into account, the quantum Langevin equations (QLEs) derived from the Hamiltonian (\ref{eq:paramH}) are
\begin{align}\label{eq:QLE}
\partial_t a &= -\frac{\kappa}{2} a + 2 g a\dg b + i \eta_A, \nonumber \\
\partial_t b &= - \frac{\Gamma}{2} b - g a^2 + \mathcal{E} + i \eta_B,
\end{align}
where $\kappa$ ($\Gamma$) is the photonic (phononic) decay rate and the $\eta_i$s are stochastic Gaussian noise operators. Since the QLEs (\ref{eq:QLE}) are nonlinear, there is no generic way to solve these equations. As the deterministic part of the steady-state QLEs is invariant under the photonic parity transformation $a \rightarrow -a$, we expect $\expval{a}_{\text{ss}} = 0$. This point will be addressed in Section IV. As a start in understanding the behavior of the system, one can find semiclassical steady-state solutions to the QLEs (\ref{eq:QLE}). These are 
\be\label{eq:meanfieldsteadystate}
(\alpha, \beta)_{\text{ss}} = \left( 0, \frac{2 \mathcal{E}}{\Gamma} \right), \hspace{4pt} (\alpha, \beta)_{\text{ss}} = \left(\pm \sqrt{\frac{\mathcal{E} - \mathcal{E}_c}{g}}, \frac{\kappa}{4g} \right),
\ee
where $\mathcal{E}_c = \kappa \Gamma/(8 g)$ is the critical driving. Importantly, the second solution above only exists when $\mathcal{E} > \mathcal{E}_c$, so that we may define an above-threshold and a below-threshold phase. Below threshold, the photonic field amplitude equals zero, whereas the mechanical amplitude increases linearly with the driving. Above threshold, the mechanical amplitude saturates, an effect known as pump depletion \cite{Charmichaelbook1, MilburnWallsbook}, and all the energy pumped into the system is converted to the subharmonic (photonic) mode. As the rate at which this happens scales with $\mathcal{E}$, the photonic amplitude grows as $\sqrt{\mathcal{E}}$. The emergent picture is that of a phase transition. Because of this, a simple linearizaton procedure $a \rightarrow \alpha_{\text{ss}} + a$, $b \rightarrow \beta_{\text{ss}} + b$ yields diverging results for the system's observables near the threshold $\mathcal{E}_c$. A way to overcome this problem is the use of a self-consistent linearization \cite{Veits}.

It is more convenient to write the QLEs (\ref{eq:QLE}) in terms of the scaled variables $\epsilon = \mathcal{E}/\mathcal{E}_c$, $x=\kappa \Gamma/(8 g^2)$, $y=\kappa^2/(16g^2)$, $a=\sqrt{x}\,\tilde{a}$, $b=\sqrt{y}\,\tilde{b}$, $\eta_A = \sqrt{\kappa}\, \xi_A$, and $\eta_B = \sqrt{\Gamma}\, \xi_B$. The inverse of $x$, is related to the single-photon cooperativity $4g^2/(\kappa \Gamma)$. With this scaling, the critical point is fixed at $\epsilon_c =1$, thus facilitating the comparison of solutions with different parameters, and the quantitative comparison of the stochastic fluctuations in the diabatic regime ($\Gamma \ll \kappa$). With the scaled variables, the QLEs (\ref{eq:QLE}) reduce to
\begin{align}\label{eq:scaledQLE}
\gamma \partial_\tau \tilde{a} &= - \tilde{a} + \tilde{a}\dg \tilde{b} +i \sqrt{\frac{2}{x}} \xi_A (\tau/\gamma), \nonumber \\
\partial_\tau \tilde{b} &= -\tilde{b} -  \tilde{a}^2 +\epsilon +i \sqrt{\frac{2}{y}} \xi_B (\tau),
\end{align}
where $\tau = \frac{1}{2}\Gamma t$, $\gamma = \Gamma / \kappa$ and $\expval{\xi_i(t), \xi_i\dg(t')} = (\bar{n}_i+1)\delta(t-t')$.  Here and onwards, we  consider that there are no thermal fluctuations for the photonic mode ($\bar{n}_A=0$), and that the system is in the diabatic limit, where the phonon dissipation is negligible, $\gamma \ll 1$. As the smallest $y$ for microwave resonators\cite{Aspelmeyer} is $\sim 10^{5}$, the contribution of $\xi_B$ is negligible for low thermal phonon numbers ($\bar{n}_B \ll y$). Another noise contribution to the dynamics of the mechanical mode $\tilde{b}$ originates from the fluctuations of the coupling term: $\tilde{a}\tilde{a} - \expval{\tilde{a} \tilde{a}}$. These fluctuations are proportional to the quantum noise introduced by $\xi_A$ and occur on the much faster timescale $\tau/\gamma$ (they scale with $\langle\xi_A \xi_A^{\dagger}\rangle =\gamma \delta ( \tau - \tau')$). Thus, in the diabatic limit, these fluctuations vanish as well.

Having discussed the fluctuations of the scaled QLEs (\ref{eq:scaledQLE}), we turn to the self-consistent linearization procedure. We proceed by linearizing the phononic field around the steady-state amplitude 
\be\label{eq:implicit}
\tilde{\beta}_{\text{ss}} = \epsilon - \tilde{\alpha}_{\text{ss}}^2,
\ee
where $\tilde{\alpha}_{\text{ss}}$ is still to be determined. We let $b \to \tilde{\beta}_{ss} + \delta b$ and substitute this result into the QLEs (\ref{eq:QLE}). Throwing away terms with products of fluctuations leads to
\be\label{eq:linQLE}
\gamma \partial_\tau \tilde{a} = -\tilde{a} + \tilde{\beta}_{\text{ss}} \tilde{a}\dg + i \sqrt{\frac{2}{x}} \xi_A,
\ee
which can be readily solved for the photonic field $\tilde{a}$. Solving Eq.(\ref{eq:linQLE}) enables the evaluation of all steady-state photonic moments $\langle\tilde{a}^{\dagger n} \tilde{a}^m\rangle$. Importantly, one finds
\begin{align}\label{eq:moments}
\expval{\tilde{a}\dg \tilde{a}}_{\text{ss}} =\tilde{\beta}_{\text{ss}}\expval{\tilde{a}^2}_{\text{ss}} =  \frac{\tilde{\beta}^2_{\text{ss}}}{2x(1-\tilde{\beta}_{\text{ss}}^2)}, 
\end{align}
where $\tilde{\beta}_{\text{ss}}$ still needs to be determined self-consistently. Combining Eq. (\ref{eq:moments}) with Eq. (\ref{eq:implicit}), one arrives at
\be\label{eq:selfconsisrel}
\tilde{\beta}_{\text{ss}} = \epsilon - \frac{\tilde{\beta}_{\text{ss}}}{2x(1-\tilde{\beta}_{\text{ss}}^2)}.
\ee
The solution to Eq. (\ref{eq:selfconsisrel}) is
\begin{align}\label{eq:beta_self-consis}
\tilde{\beta}_{ss} &= \frac{\epsilon}{3} -\frac{\left(1-i \sqrt{3}\right) \aleph}{6 \sqrt[3]{\epsilon\, \beth+i\sqrt{\aleph^3 -\epsilon^2\beth^2} }} \nonumber \\
& -\frac{\left(1+i \sqrt{3}\right)}{6} \sqrt[3]{\epsilon \,\beth+i \sqrt{\aleph^3 -\epsilon^2\beth^2} }\,,
\end{align}
where 
\be
\aleph=\epsilon^2+\frac{3}{2x}+3\quad \text{and} \quad \beth=\epsilon^2 +\frac{9}{4x}-9\,.
\ee

\begin{figure}
\centering
\includegraphics[scale=0.49]{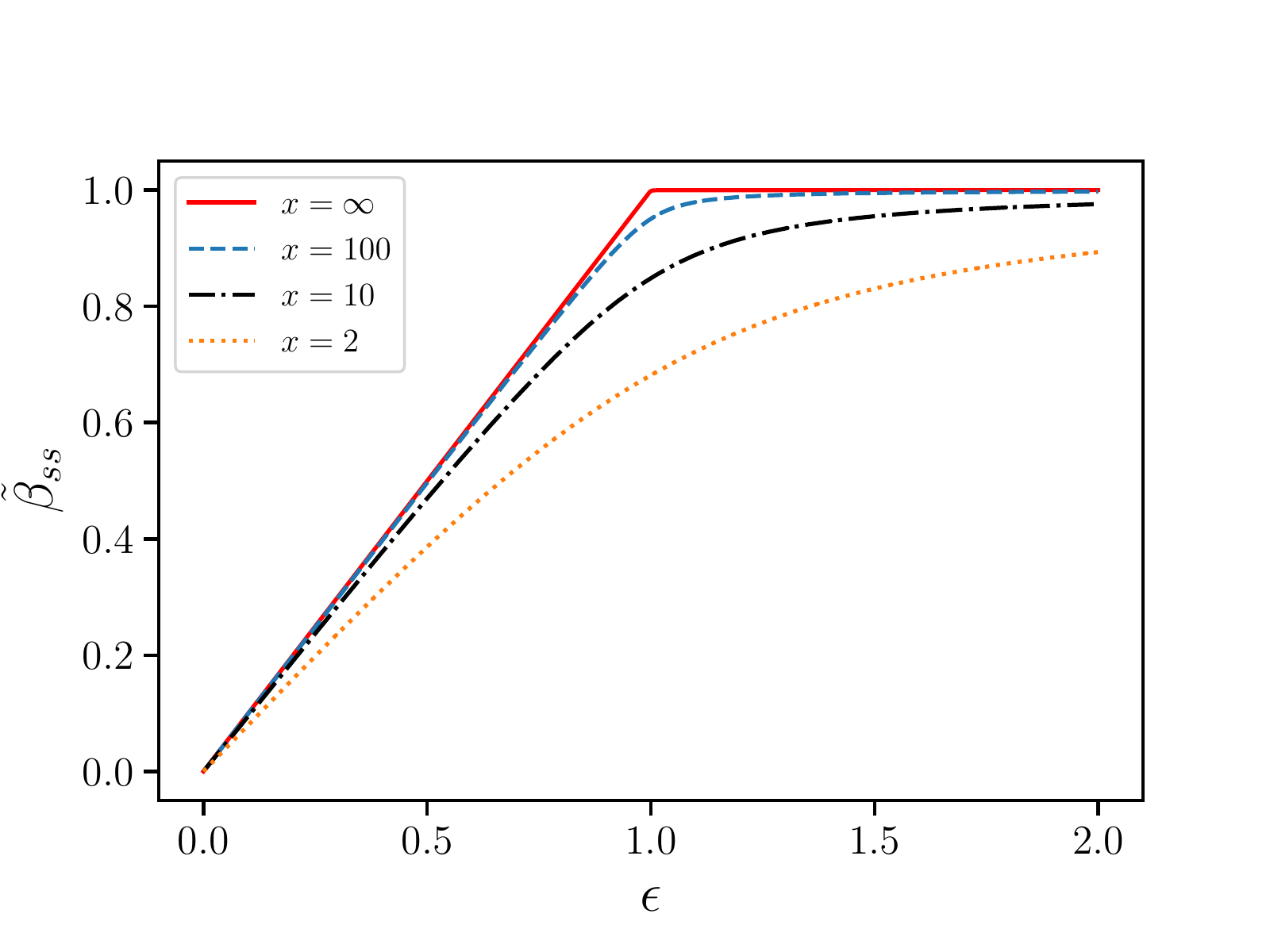}
\caption{Scaled steady-state mechanical amplitude as a function of driving. For lower $x$, the amplitude exhibits a more rounded transition near the threshold. When the single-photon cooperativity approaches zero, the noiseless classical result is reproduced.}
\label{fig:scaled-plots-A}
\end{figure}
Fig. \ref{fig:scaled-plots-A} shows the normalized mechanical amplitude $\tilde{\beta}_\text{ss}$ as a function of the normalized driving $\epsilon$. For large $x$, when the coupling is small with respect to the geometric average of the dissipation rates, the self-consistent result tends towards the semiclassical solution. This happens because the fluctuations introduced by the noise operator are suppressed (see Eq. (\ref{eq:scaledQLE})). For lower $x$, the self-consistent result deviates significantly from the semiclassical prediction near threshold. In this case, the noise introduced by $\xi_A$ plays a significant role. In general, we expect the self-consistent approach to yield exact results in the limit $y \rightarrow \infty$, where the fluctuations of the mechanical oscillator play no role.

\begin{figure}
\includegraphics[scale=0.49]{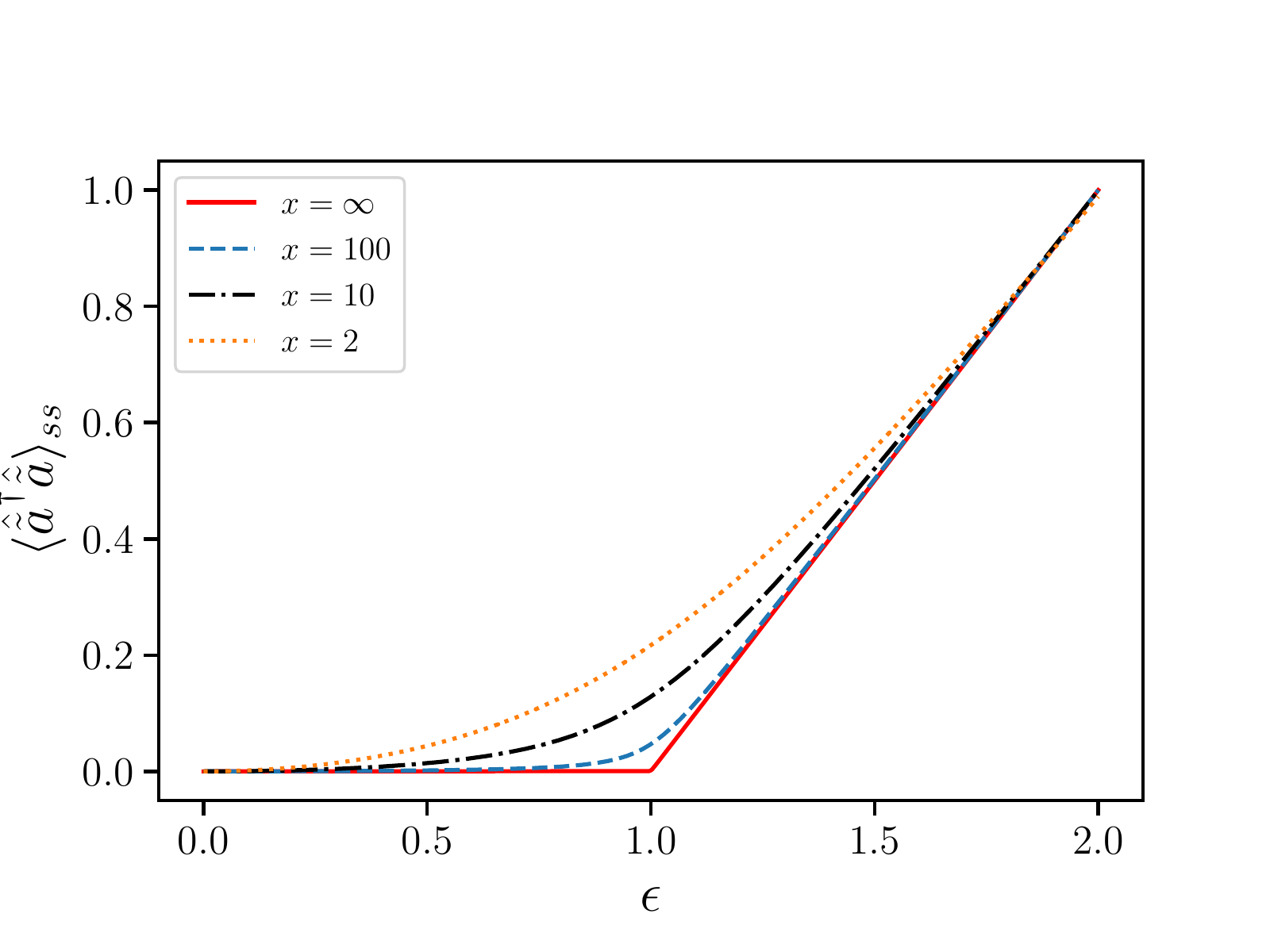}
\caption{Scaled steady-state photon number as function of driving. For lower $x$, the system exhibits a smoother (more rounded) transition near the threshold. The semiclassical result is met only for high and low driving.}
\label{fig:scaled-plots-B}
\end{figure}
Plugging Eq.(\ref{eq:beta_self-consis}) into Eq. (\ref{eq:moments}), one arrives at the self-consistent linearized expressions for the moments. Fig. \ref{fig:scaled-plots-B} shows the normalized expected photon number $\expval{ \tilde{a}\dg \tilde{a} }$ as a function of the normalized driving $\epsilon$. Again, for large $x$, the self-consistent result tends to the semiclassical prediction. For smaller $x$, it deviates from the semiclassical prediction because the fluctuations are no longer negligible. The most important difference from the semiclassical picture is that microwave photons are still created below threshold, due to zero-point fluctuations.

\section{The Fokker-Planck approach in the complex P-representation}
Although the self-consistent linearization method succeeds in producing analytical results, the results still differ qualitatively from numerical results (see Section IV). In order to obtain better results, we resort to the technique of FP equations in the complex P-representation. To derive a FP equation describing the degenerate parametric oscillator,  it is convenient to start from a master equation for the density matrix of the Lindblad type:
\begin{align}\label{eq:Lindblad}
d_t \rho_S = &-i \left[H_S, \rho_S \right]  + \kappa \mathcal{L} ( a ) \rho_S + \Gamma  \bar{n}_{B} \mathcal{L} (b^{\dagger}) \rho_S \nonumber \\
&+  \Gamma  \left( \bar{n}_{B} + 1 \right) \mathcal{L} (b) \rho_S ,
\end{align}
where $H_S$ is given in Eq. (\ref{eq:paramH}), and the superoperator $\mathcal{L}(\mathcal{O})$ is defined as
\be
\mathcal{L} (\mathcal{O}) \rho_S \equiv \mathcal{O} \rho_S \mathcal{O}^{\dagger} - \frac{1}{2} \left( \mathcal{O}^{\dagger} \mathcal{O} \rho_S + \rho_S \mathcal{O}^{\dagger} \mathcal{O}   \right).
\ee
The path from the Lindblad equation (\ref{eq:Lindblad}) to the FP equation in the complex $P$-representation \cite{Drummondgen} is well-trodden so that here we only present the result \cite{Charmichaelbook1, MilburnWallsbook}:
\begin{align}\label{eq:fullFP}
\partial_t&  P \left(  \Theta, t \right)  = \bigg\{ \partial_{\alpha_1} \left[ \frac{\kappa}{2} \alpha_1 - 2g \beta_1 \alpha_2 \right] + \partial_{\alpha_2} \left[ \frac{\kappa}{2} \alpha_2 - 2g \beta_2  \alpha_1 \right] \nonumber \\
&+   \partial_{\beta_1} \left[ \frac{\Gamma}{2} \beta_1 + g \alpha_1^2 + \mathcal{E} \right] + \partial_{\beta_2} \left[ \frac{\Gamma}{2} \beta_2 + g \alpha_2^2 + \mathcal{E} \right] \nonumber \\
&+g \left[ \partial_{\alpha_1}^2 \beta_1 \right.  + \left. \partial_{\alpha_2}^2 \beta_2 \right] + \Gamma \bar{n}_{B} \partial_{\beta_1} \partial_{\beta_2} \bigg\} P \left(  \Theta, t \right),
\end{align}
where $\Theta = (\alpha_1, \alpha_2, \beta_1, \beta_2)$. Using the same scaled variables as before, the FP equation becomes
\begin{align}\label{eq:fullFPextra}
&\partial_\tau P\left( \tilde{\Theta}, \tau \right) = \bigg\{ \frac{1}{\gamma}\partial_{\tilde{\alpha}_1} \left[\tilde{\alpha}_1 -\tilde{\beta}_1 \tilde{\alpha}_2 \right] + \frac{1}{\gamma}\partial_{\tilde{\alpha}_2} \left[\tilde{\alpha}_2 - \tilde{\beta}_2  \tilde{\alpha}_1 \right] \nonumber \\ &+   \partial_{\tilde{\beta}_1} \left[\tilde{\beta}_1 +\tilde{\alpha}_1^2 + \epsilon \right] + \partial_{\tilde{\beta}_2} \left[\tilde{\beta}_2 + \tilde{\alpha}_2^2 + \epsilon \right] \nonumber \\
&+ \frac{1}{2\gamma x} \left[ \partial_{\tilde{\alpha}_1}^2\tilde{\beta}_1 +\partial_{\tilde{\alpha}_2}^2 \tilde{\beta}_2 \right] + \frac{2\bar{n}_{B}}{y} \partial_{\tilde{\beta}_1} \partial_{\tilde{\beta}_2} \bigg\} P\left( \tilde{\Theta}, \tau \right).
\end{align}
Eq.(\ref{eq:fullFPextra}) is not easily solvable, and in order to make progress it must be brought to a simpler form. As before, the mechanical mode can be eliminated in the steady-state and the thermal noise of the mechanical oscillator can be disregarded. The validity of this approach can be seen from the stochastic differential equations (SDEs) corresponding to Eq. (\ref{eq:fullFPextra}). Using the It\^{o} rules, the SDEs are found to be \cite{Drummond1981}:
\begin{align}\label{eq:SDE}
\gamma\partial_\tau \begin{bmatrix}
\tilde{\alpha}_1 \\ \tilde{\alpha}_2
\end{bmatrix} & = \begin{bmatrix}
\tilde{\alpha}_2 \tilde{\beta}_1 - \tilde{\alpha}_1 \\ \tilde{\alpha}_1 \tilde{\beta}_2 - \tilde{\alpha}_2 
\end{bmatrix} + \begin{bmatrix}
\frac{\tilde{\beta}_1}{2x} & 0 \\
0 & \frac{\tilde{\beta}_2}{2x}
\end{bmatrix}^{\frac{1}{2}} \begin{bmatrix}
\zeta_{A_1} \\ \zeta_{A_2}
\end{bmatrix}, \nonumber \\
\partial_\tau \begin{bmatrix}
\tilde{\beta}_1 \\ \tilde{\beta}_2
\end{bmatrix} & = \begin{bmatrix}
\epsilon - \tilde{\alpha}_1^2 - \tilde{\beta}_1 \\ \epsilon - \tilde{\alpha}_2^2 - \tilde{\beta}_2
\end{bmatrix} + \begin{bmatrix}
 0& \frac{2\bar{n}_{B}}{y} \\
\frac{2\bar{n}_{B}}{y}& 0
\end{bmatrix}^{\frac{1}{2}} \begin{bmatrix}
\zeta_{B_1} \\ \zeta_{B_2}
\end{bmatrix},
\end{align}
where $\expval{\zeta_{i_1}} = \expval{\zeta_{i_2}} = 0$ and $\expval{\zeta_{i_1} \zeta_{i_2}} = \delta (t-t')$.

In the adiabatic regime ($\Gamma \gg \kappa$), these coupled SDEs can be reduced to a simpler form by substituting in the steady-state solution $\tilde{\beta}_{\text{ss},i} = \epsilon - \tilde{\alpha}_{\text{ss},i}^2$ in the upper two equations of relation (\ref{eq:SDE}). This adiabatic elimination is especially convenient for low phonon thermal numbers ($\bar{n}_B \ll y$), so that the thermal fluctuations of the mechanical oscillator can be disregarded. This is the typical case in quantum optics and the resulting steady-state FP equation was already solved analytically \cite{Drummond1981, MilburnWallsbook}. The situation for higher $\bar{n}_\beta $ has also been considered \cite{Drummond1981}. 

For electromechanical systems, however, one is typically concerned with the diabatic regime ($\Gamma \ll \kappa$), where adiabatic elimination is not valid.  However, for the steady-state of the system, it is still possible to derive the same FP equation for the diabatic regime, and compute observables to very good approximation. 

\begin{figure}
\includegraphics[scale=0.5]{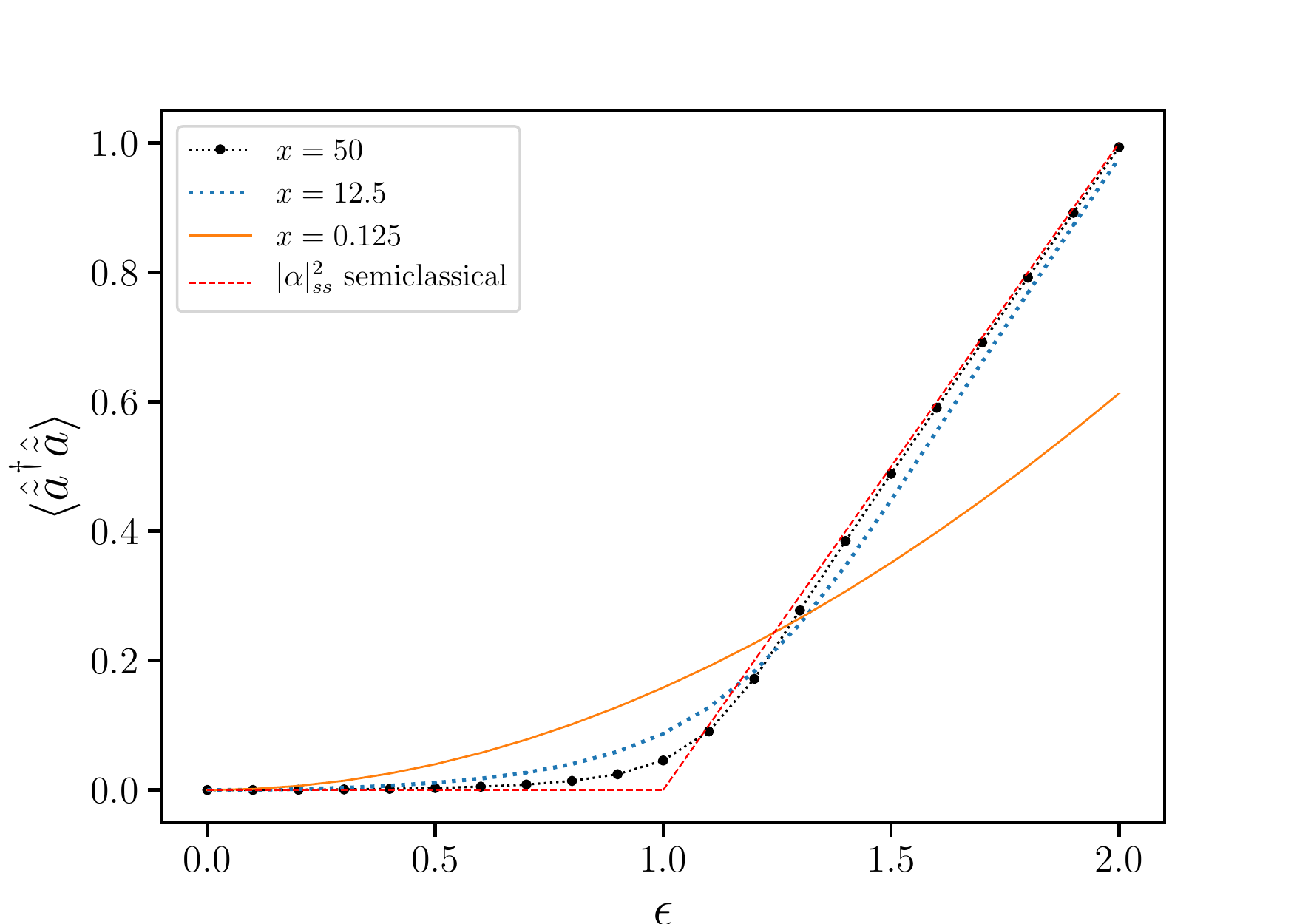}
\caption{Steady-state photon number as a function of driving for different values of $x$. For $x=50$, the FP approach converges to the semiclassical prediction as the fluctuations of the mechanical oscillator become small. For $x = 12.5$, the scaled photon number "undershoots" the semiclassical prediction just above threshold. This behavior is not predicted in the linearized approach. For $x=0.125$, the quantum fluctuations introduced by $\xi_A$ lead to a more pronounced undershooting.}
\label{fig:phot}
\end{figure}

Outside the adiabatic regime, it is in general possible to consider the substitution of the steady-state solution $\tilde{\beta}_{\text{ss,i}}$ as a linearization around the deterministic steady-state: $\tilde{\beta}_i = \tilde{\beta}_{\text{ss},i} + \delta \tilde{\beta}_i$, where $\tilde{\beta}_{\text{ss},i} = \epsilon - \tilde{\alpha}_{\text{ss},i}^2$ and $\delta \tilde{\beta}_i$ represents the fluctuations around the steady-state. From Eqs. (\ref{eq:scaledQLE}), we know that in the diabatic regime, when $\kappa \gg g \gg \Gamma$, the thermal fluctuations affecting the mechanical resonator become small, close to the steady-state. Thus, we expect the linearization to produce good results as long as the fluctuations are small.
Substituting the steady-steady solution $\tilde{\beta}_{\text{ss},i}$ in the SDEs (\ref{eq:SDE}), these become 
\begin{align}\label{eq:redSDE}
\gamma\partial_\tau \begin{bmatrix}
\tilde{\alpha}_1 \\ \tilde{\alpha}_2
\end{bmatrix} & = \begin{bmatrix}
\tilde{\alpha}_2 \tilde{\beta}_{\text{ss,1}} - \tilde{\alpha}_1 \\ \tilde{\alpha}_1 \tilde{\beta}_{\text{ss,2}} - \tilde{\alpha}_2 
\end{bmatrix} + \begin{bmatrix}
\frac{\tilde{\beta}_{\text{ss,1}}}{2x} & 0 \\
0 & \frac{\tilde{\beta}_{\text{ss,2}}}{2x}
\end{bmatrix}^{\frac{1}{2}} \begin{bmatrix}
\zeta_{A_1} \\ \zeta_{A_2}
\end{bmatrix}.
\end{align}
The steady-state of the FP equation associated with Eq.(\ref{eq:SDE}) is
\begin{align}\label{eq:effectiveFP}
&0 = \bigg\{ \partial_{\tilde{\alpha}_1} \left[\tilde{\alpha}_1 -\left( \epsilon - \tilde{\alpha}_{1}^2 \right)  \tilde{\alpha}_2 \right] + \partial_{\tilde{\alpha}_2} \Big[ \tilde{\alpha}_2 - \left( \epsilon - \tilde{\alpha}_{2}^2 \right) \tilde{\alpha}_1 \Big] \nonumber \\
& +\frac{1}{2x} \Big[ \partial_{\tilde{\alpha}_1}^2 \left( \epsilon - \tilde{\alpha}_{1}^2 \right)+ \partial_{\tilde{\alpha}_2}^2 \left( \epsilon - \tilde{\alpha}_{2} ^2 \right) \Big]\bigg\} P_\text{ss}(\tilde{\alpha}_1, \tilde{\alpha}_2).
\end{align}
This partial differential equation is equivalent to that studied in the quantum optics situation \cite{Drummond1981, MilburnWallsbook}. The solution to Eq.(\ref{eq:effectiveFP}) is readily found to be
\begin{equation}\label{eq:Pfunc}
P_{\text{ss}} = \mathcal{N}\left[ \left(\epsilon - \tilde{\alpha}_1^2\right)\left(\epsilon - \tilde{\alpha}_2^2\right)\right]^{x-1}e^{2 \tilde{\alpha}_1 \tilde{\alpha}_2x},
\end{equation}
where $\mathcal{N}$ is a normalization constant.
The photonic moments are defined as \cite{Drummondgen}
\begin{equation}\label{eq:definitionmoments}
\expval{\tilde{a}^{\dagger n} \tilde{a}^m} = \int_{\mathcal{C}} \int_{\mathcal{C'}} d \tilde{\alpha}_1 d \tilde{\alpha}_2 \tilde{\alpha}_2^n \tilde{\alpha}_1^m P(\tilde{\alpha}_1, \tilde{\alpha}_2).
\end{equation}
For the complex $P$-distribution (\ref{eq:Pfunc}) an analytical expression for arbitrary $n,m$ can be written as
\begin{align}\label{eq:momentspecific}
&\expval{\tilde{a}^{\dagger n} \tilde{a}^m}_{\text{ss}} =  \mathcal{N}' \sum_{k=0}^{\infty} \frac{(2x)^k}{k!} \left( \sqrt{\epsilon} \right)^{k+m} \left( \sqrt{\epsilon} \right)^{k+n} \nonumber \\
& \times \prescript{}{2}F_1(-(k+m),x,2x,2)\prescript{}{2}F_1(-(k+n),x,2x,2),
\end{align}
where $\prescript{}2F_1(u,v,w,z)$ is the hypergeometric function and $\mathcal{N}'$ is a normalization constant. It is also possible to obtain an analytical expression for the steady-state mechanical amplitude as it is defined in terms of the second order moment $\expval{a^2}_{\text{ss}}$. Fig. \ref{fig:phot} shows the steady-state photonic field intensity as function of the driving for different values of $x$. The FP approach predicts an "undershooting" of the semiclassical prediction just above threshold, which becomes more pronounced when $x$ is small; that is, when the coupling is large with respect to the dissipation rates. For $x=50$, the FP approach tends towards the semiclassical prediction as the fluctuations of the mechanical and LC oscillator become small. For $x = 12.5$, the undershooting is relatively small because the quantum fluctuations introduced to $\tilde{a}$ by $\xi_A$ in Eq. (\ref{eq:scaledQLE}) are suppressed. In contrast, for $x = 0.125$, the quantum fluctuations are relatively large, and the undershooting is more pronounced.  

\section{Comparison of Approaches with Numerical Results}

To check the performance of the FP approach outside the adiabatic regime, we compare our results with numerical simulations. The simulations are performed for $\kappa = \Gamma =1$, $g=0.1$, where the system is still tractable numerically, and are done using QuTiP \cite{qutip}. To obtain precise results for the observables, the dimension of the matrix representing the operator should be significantly bigger than the expected outcome of the observable's
value. Constructing a Fock-state basis containing $2N$ photons and $N$ phonons requires all matrices to have size $2N^2 \times 2N^2$. We found that taking N=23 provides reasonable precision while avoiding memory issues. However, for higher values of the driving, this dimension is not sufficient. In these cases, we used Shanks-extrapolation with as input the results of $N=15,\ldots,23$ to improve the results. The chosen region of parameter space lies neither in the adiabatic nor in the diabatic regime, so that the FP approach can at best be regarded as a linearization. As $x = 12.5$ and $y = 6.25$, we expect the noise contributions $\xi_A$ and $\xi_B$ to be negligible, making this linearization a reasonable approximation in the steady-state. 

\begin{figure}
\includegraphics[scale=0.49]{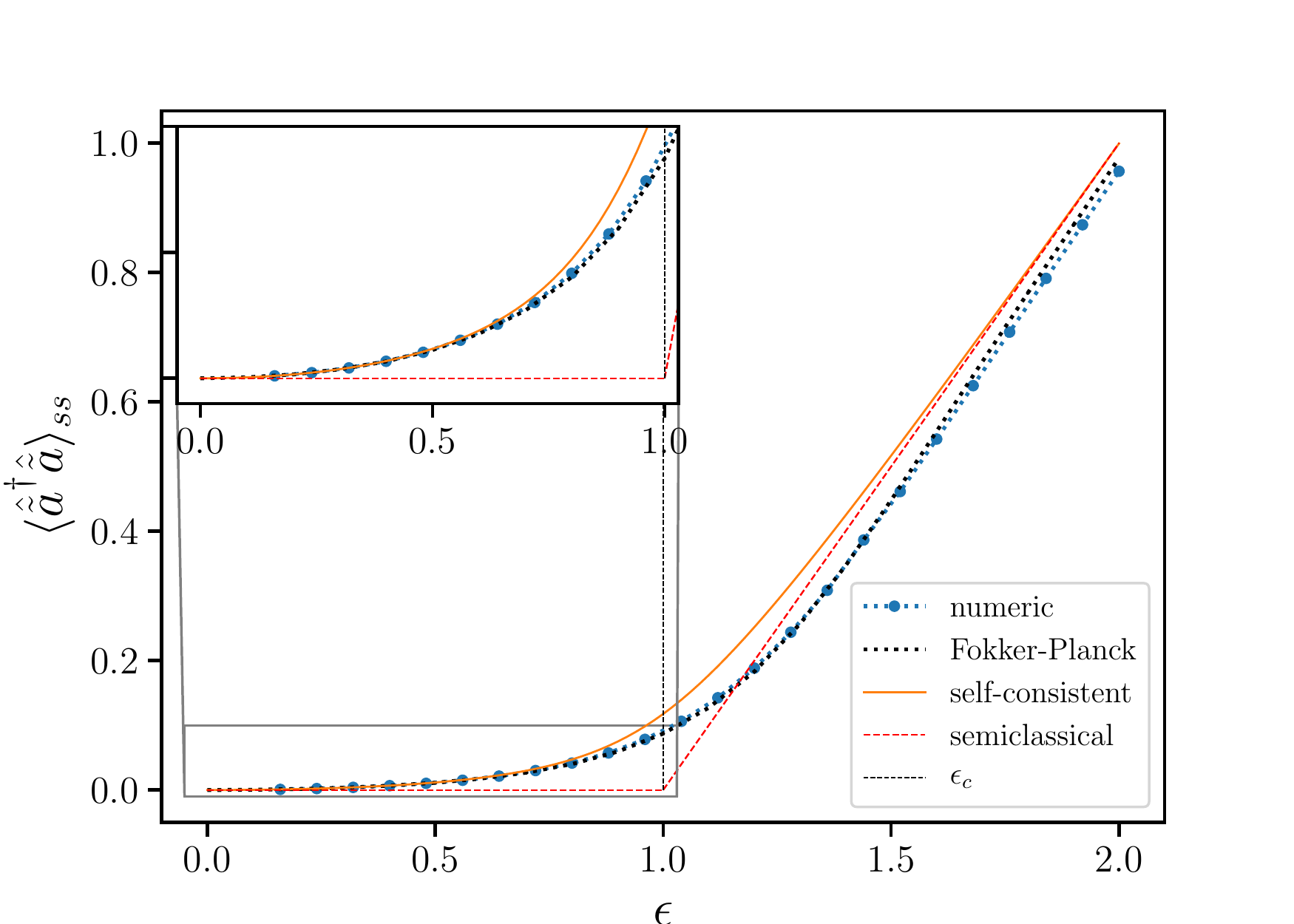}
\caption{Normalized photon number as a function of the normalized driving. The FP approach corresponds very well with the numerical data.}
\label{fig:phot_num-ss}
\end{figure}
\begin{figure}
\includegraphics[scale=0.49]{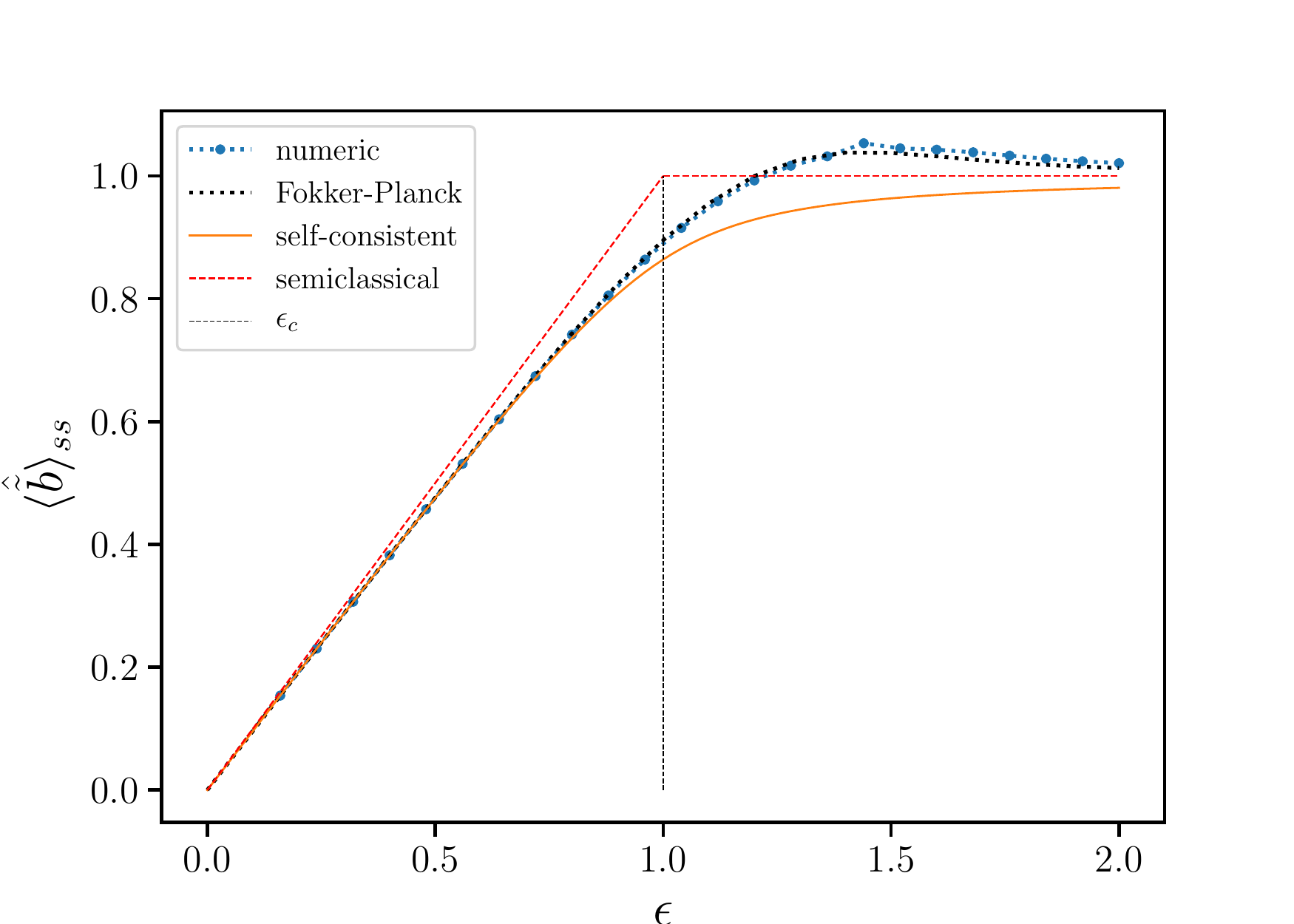}
\caption{Normalized mechanical amplitude as a function of the normalized driving. Numerical simulations show that the actual amplitude overshoots the self-consistent prediction in the intermediate driving regime.}
\label{fig:beta-ss}
\end{figure}

Figs. \ref{fig:phot_num-ss} and \ref{fig:beta-ss} show respectively the scaled steady-state photon number and mechanical amplitude
obtained via different methods. The numerical result is shown in dotted-blue. The dashed-red line represents the semiclassical (mean-field) solution to the steady-state QLEs (\ref{eq:meanfieldsteadystate}). The numeric behavior of the system coincides roughly with the semiclassical result, but it deviates from it around the threshold. The self-consistent linearization is shown in solid-orange, and it coincides with the semiclassical result for high and low
driving power but connects these limits smoothly (in contrast with the sharp transition of the semiclassical result).
However, it does not predict the undershooting of the photon number and the overshooting of the mechanical amplitude just above threshold, unlike the numerical predictions. The FP approach (dotted-black) does predict the overshooting and undershooting and it is qualitatively similar to the numeric result. Upon close inspection, it is seen that the numerical results reaches the semiclassical result for higher driving slightly faster than the FP result. This small discrepancy can have two origins. Firstly, the numerical simulations become less precise for higher values of the driving. Secondly, in the $\kappa=\Gamma$ regime, the FP method gives a linearized approximation of the true steady-state observables implying that the discrepancy could originate in shortcomings of the analytical method.

\begin{figure}
\centering
\includegraphics[scale=0.28]{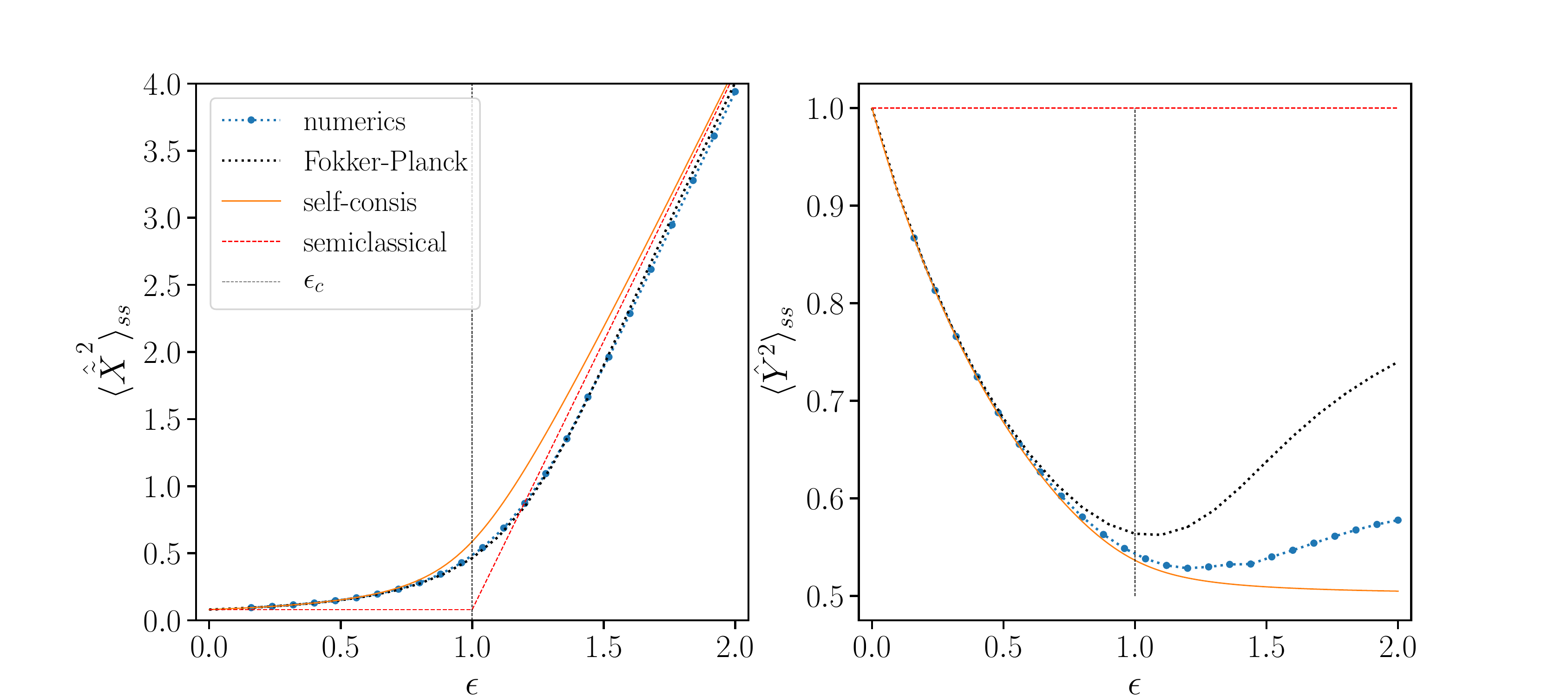}
\caption{Steady-state fluctuations of the scaled photonic $\tilde{X}$- and unscaled $Y$-quadratures as a function of driving. The behavior of $\langle \tilde{X}^2\rangle$ is qualitatively similar to that of the photon number. For the $Y$-quadrature, the fluctuations decrease below 1 with driving, indicating the presence of squeezing. A yet unexplained discrepancy is observed between the numerical data, the self-consistent and the FP approaches.}
\label{fig:quadratures-ss}
\end{figure}

Fig. \ref{fig:quadratures-ss} shows the steady-state fluctuations of the photonic quadratures $X = a\dg + a$ and $Y = i(a\dg - a)$ for the different methods. The behavior of the fluctuations for the $\tilde{X}$-quadrature is qualitatively similar to that of the photon number. The fluctuations for the $Y$-quadrature drop below $1$ for nonzero driving, suggesting that the photonic state is squeezed. To clarify this, the plot shows the unscaled quadrature $Y$. Above threshold, a clear discrepancy between the FP result and the self-consistent result can be seen. This discrepancy has been reported before \cite{Navarrete-Benlloch}, but has not yet been explained. The numerical result does not coincide with either one. This might be caused by the dimensional issues discussed above.

\begin{figure}
\centering
\includegraphics[scale=0.4]{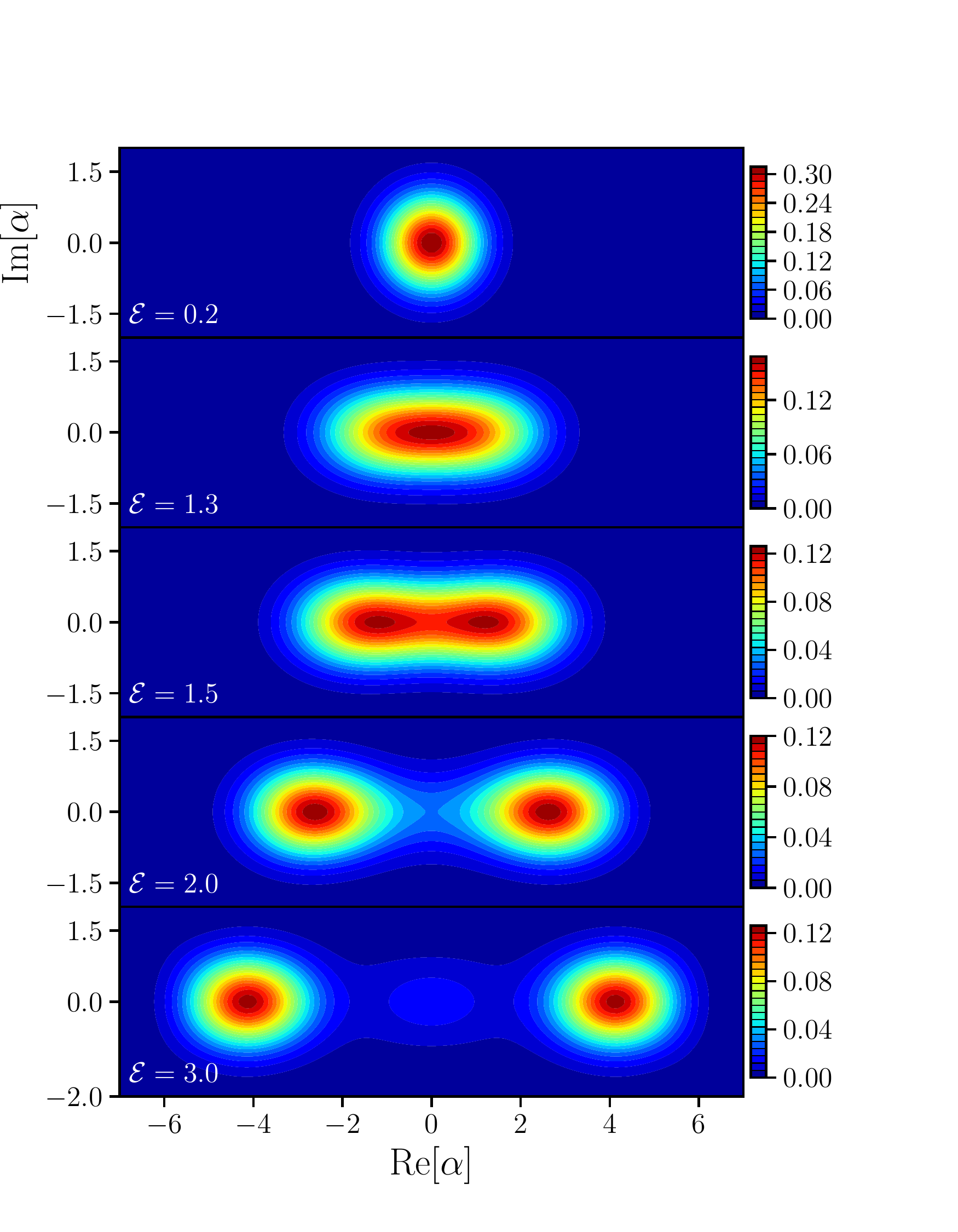}
\caption{Q-function of the LC steady-state for different values of driving. For well-resolved density maxima, as for $\mathcal{E} = 3.0$, the centers of the maxima lie at the branches of the square root defined in Eq. \ref{eq:meanfieldsteadystate}.}
\label{fig:superposition-alpha-ss}
\end{figure}

As a final note, we come back to the invariance of the mean-field steady-state equations of motion with respect to the transformation $\alpha \rightarrow -\alpha$. In Section II, we noted that this implies that $\alpha_\text{ss} =0$, which is indeed what follows from computing it through Eq. (\ref{eq:momentspecific}). For this to be consistent with the phase transition behavior obtained via the semiclassical approximation, the system must go from the Gaussian vacuum state into a mixture of coherent states near and above threshold. This behavior is confirmed by numerical simulations, as seen from the evolution of the cavity state with the driving, as displayed in Fig. \ref{fig:superposition-alpha-ss}.

\par

\section{Conclusion}
We have discussed the behavior of an electromechanical system in the parametric regime, where $2 \omega_{LC} \sim \Omega_m$. In this regime and with RWA, the Hamiltonian effectively assumes the form of the degenerate parametric oscillator well-known in quantum optics. It is known from the latter's context, that the system undergoes a phase transition: below the critical driving, the mechanical amplitude grows linearly with the driving while the photon number is unaffected (in a semiclassical picture); above the critical driving, the mechanical amplitude saturates and the photon number grows with the driving. For the electromechanical case, this transition takes place when the phonon number created by the mechanical driving reaches the value $(8g/\kappa)^{-2}$, which requires at least $\sim 10^{4}$ pump phonons for current devices \cite{Aspelmeyer}. We also find for all the different quantum approaches that quantum fluctuations smear this transition, and that photons can be created below the threshold.

In contrast with the quantum optical situation, the mechanical dissipation rate in electromechanical systems is usually much smaller than the photonic dissipation rate. However, in this diabatic limit, the fluctuations of the mechanical mode become negligible \cite{Veits}. We have shown that in this case, it is possible to linearize the SDEs corresponding to the full Fokker-Planck equation to arrive at the same effective SDEs that describe the quantum optical system after adiabatic elimination. In the diabatic limit, where $\gamma \rightarrow 0$, and in steady-state, this method provides a very good approximation, and known analytical results \cite{Drummond1981} can be extended. With these results, an expression for the steady-state mechanical amplitude can be found self-consistently.

 We find that the use of a Fokker-Planck approach agrees better with the numerical results (specifically for the parameter values $\kappa = \Gamma =1$, $g=0.1$), and in contrast with the standard linearization techniques, the FP approach and the numerical simulations both predict an undershooting of the photon number above threshold. We were not able to check whether the FP method provides close-to-exact analytical results in the diabatic limit for the steady-state moments. 

Although the focus of this paper is the parametric regime in electromechanics, it might be possible to find analogous effects in other optomechanical systems, such as a membrane-in-the-middle with quadratic coupling driven at $\Delta=-2\Omega$.
Whereas experiments on electromechanical systems in the parametric regime have not yet been reported in the literature, they can be performed with the existing technology.

\textit{Acknowledgements}- We thank the Dutch Science Foundation (NWO/FOM) for its financial support.\\

\bibliographystyle{ieeetr}

\end{document}